# (World) Building Transformation: Students and Teachers as Co-Creators in OpenXR Learning Environments

Greenbaum, A. , Strickler, E. , Patterson, V. , & Oluleye, B.

Extended Reality    human-centered design thinking    metaverse

problem-posing education    Transformative Learning

XR learning environments

*Emerging extended reality (XR) tools and platforms offer an exciting opportunity to align learning experiences in higher education with the futures in which students will pursue their goals. However, the dynamic nature of XR as subject matter challenges hierarchies and classroom practices typical of higher education. This instructional design practice paper reflects on how our team of faculty, learning experience designers, and user experience (UX) researchers implemented human-*

*centered design thinking, transformative learning, and problem-posing education to design and implement a special topics media entrepreneurship course in building the metaverse. By pairing our practitioner experience with learner personas, as well as survey, interview, and focus group responses from our learners, we narrate our design and its implications through a human-centered, reflective lens.*

# Introduction

Consider this learner. She is a Media Entrepreneurship major interested in fashion design, and she has been thinking about expanding her projects into virtual worlds like she has seen her friends do. "I want to create more in the virtual world," she will tell you. "It will bring a whole new outlook on how we create and interact with each other." She spends time online with an international friend group in Roblox, a pioneering online platform that enables users to generate, share, and experience a variety of interactive virtual worlds.

Also, consider the learner who is aware of his own limits. "I'm quite slow when it comes to these new technologies," he perceives. "It makes me less likely to pursue going in depth with them." He experiences anxiety about keeping up with his studies and the technologies that will matter in his career – he wonders if there will always be something new that he doesn't know how to use and what the learning curve will entail.

We introduce these student personas in this instructional design practice paper because our design thinking approach begins by seeking insight from people most impacted by design decisions. Both student personas perceive emerging technologies as an area with significant potential to broaden their horizons and enhance their career. They are interested in building skills and knowledge across emerging XR landscapes. These personas informed our design process for "Self and Community in the Metaverse," the course design described in this paper.

As a convergence point of various digital trends and patterns, the metaverse offers a unique platform where students can explore the dynamics of a digitally interconnected world. The metaverse is a persistent and interconnected network of 3D virtual worlds that will eventually serve as the gateway to most online experiences and underpin much of the physical world (Ball, 2020). It is a space where students learn through play, build virtual worlds, and co-create stories. Students studying how to create and run media businesses need to learn by doing real projects in emerging business areas. Technical skills will help them understand the evolution of Web3, the next version of the Internet. Therefore, with these tools, students can become super learners, equipped to navigate and shape the metaverse and many other future trends and worlds (Grant, 2023).

Design thinking suggests that human-centered solutions develop through investigating human needs and reframing possible problems as design questions (IDEO, 2015). We defined our design questions through engagement with the design thinking process, including activity prototypes and learner tests, learner focus groups, and instructor reflection. In this project, we designed with the following questions in mind: How might we build student skills and confidence with disruptive technologies, and empower students to navigate new challenges? How might we demonstrate the relevance of technological trends to varied career goals? How might we define a supportive, accessible curriculum for an XR landscape characterized by rapidly evolving, collaborative, open-source ecosystems, and critical questions of interoperability (Ziker et al., 2021)? How might we teach equitably with technologies emerging from an industry where the students of our Minority Serving Institution are underrepresented?

This practice paper narrates and reflects on our experience as a team of faculty, instructional designers, and UX researchers in designing and implementing Self and Community in the Metaverse. We add "robustness" to our subjective reflection by including learner perspectives gathered in our design process, setting our narrative "against the frame of learner narratives" (Knowlton, 2007, p.221). As design and educational practitioners, we excavate crucial challenges and opportunities in designing, teaching, and learning XR through this reflective process.

## Defining Our Terms

- **XR** is an umbrella term that encompasses all extended virtual reality technologies, including Virtual Reality (VR), Augmented Reality (AR), and Mixed Reality (MR), which merge the physical and virtual worlds.
- **WebVR** allows users to experience VR directly in a web browser without the need to download specialized apps or the use of VR hardware like headsets.
- **Web3** refers to a new iteration of the World Wide Web based on blockchain technology, featuring decentralized apps that run on peer-to-peer networks.
- **Virtual Worlds** are computer-simulated environments where users can interact with each other and create objects within the digital space, often using avatars.
- **OpenXR** is a cross-platform standard that enables applications to access various virtual reality and augmented reality devices and platforms. It simplifies the development of immersive experiences by providing a common set of functions and concepts for XR applications and devices.

- **The metaverse** is a persistent and interconnected network of 3D virtual worlds where students learn through play, creation, and storytelling.

# Literature Review

The course design for Self and Community in the Metaverse was grounded in several instructional design principles: transformative learning, problem posing education, and human-centered design thinking (McLaughlin et al., 2022; Mezirow, 1997). Transformative learning attempts to change learners' frames of reference, establish autonomy, and encourage discourse that critiques different points of view. It shifts away from an emphasis on the acquisition of knowledge toward the reflective process of negotiating meaning through discourse (Mezirow, 1997). Learners who achieve these goals become autonomous, socially responsible individuals who succeed in today's rapidly changing society (Formenti & Hoggan-Kloubert, 2023). In a transformative learning experience, the instructor assumes the role of a facilitator or a mentor instead of an expert. The instructor's role is to foster dialogue and support learners as they redefine their perspectives (Cranton, 2016). Incorporating authentic learning activities into a transformative learning experience further deepens learners' understanding of real-world situations and encourages dialogue among learners (Ryman et al., 2009).

Virtual worlds are ideal for transformative learning and simulating potential futures, as these spaces encourage exploring new identities and experiences outside of everyday life. A key component of transformative learning is critical self-reflection and the questioning of belief systems that shape our world (Mezirow, 1997). Virtual environments can trigger intense self-reflection as the learner is confronted with an alternate manifestation of themselves via a virtual avatar or digital twin. When the learner builds an avatar, they must also identify aspects of their identity they wish to transfer into the virtual space and aspects they will leave behind (Harmon, 2011). Additionally, virtual worlds offer the learner an opportunity to explore and build environments that exist outside of the typical classroom hierarchy. These destabilizing environments can empower learners to critique existing power structures through the breakdown of traditional classroom hierarchies (Tilak et al., 2020).

Many approaches to XR learning resemble and replicate traditional classroom and university hierarchies. For instance, in the Metaversity phenomenon, students learn within a pre-existing digital twin of their university (Paykamian, 2022). In contrast, problem-posing education (Freire et al., 2014) implies that learners should be positioned as empowered creators rather than mere end-users. Therefore, instead of simply providing students with yet another technological experience as end-users, XR learners need to be encouraged to navigate, learn, use, and reflect on disruptive technologies as empowered creators. This can be achieved through Idea-Artefact Creation Pedagogy, through which students "frequently learn from a self-directed, team-based creative process, in which they analyze problems than design some kind of artifact which may or may not be a solution" (Lackéus, 2020).

Admittedly, XR course designs can be challenging due to the rapidly changing nature of the subject matter.  Rather than implementing a one-way transfer of knowledge from instructor to student, this context demanded a more intuitive and responsive model of instruction, one where the instructor facilitated and learned along with their students. Much like transformative learning encourages the instructor to guide the student (Mezirow, 1997), Paolo Freire conceptualized the one-way knowledge transfer as the banking system of education. This concept critiques education establishments, positioning students to store knowledge deposited by teachers, denying their creativity and critical consciousness, mirroring and perpetuating systems of oppression (Freire et al., 2014).

Freire's problem-posing education can also be examined and enacted through human-centered design thinking (McLaughlin et al., 2022). Human-centered design thinking seeks insight and inspiration from the people most impacted by a problem and uses collaboration to develop solutions based on their needs (IDEO, 2015). Across various industries and applications, designers use the design thinking process to identify people's needs and design responsive human-centered solutions. Due to their application across media, technology, and creative industries, design thinking activities provide an authentic framework through which media entrepreneurship students might grow their business development and innovation skills. These learning activities provide "skills and mindsets for more inclusively and resiliently addressing complex, real-world challenges" (McLaughlin et al., 2022, p.11).

# Design Context

## Project Origin

When the pandemic hit, the Creative Media Industries Institute (CMII) at Georgia State University recognized the untapped potential of VR and XR technologies in education. XR course activities have been widely identified as potential sites for classroom collaboration, accessibility, and experiential learning (Brown et al., 2020). Known for expertise in virtual production and game design, CMII hadn't yet applied these tools as a teaching medium. With the support of the Learning Design Team at Georgia State University (GSU)'s Center for Excellence in Teaching, Learning, and Online Education (CETLOE), CMII was positioned to innovate and apply our knowledge in virtual worlds.

CMII encompasses three core missions: provide training in technology, media, and arts; nurture media entrepreneurs; and partner with industry to generate research and economic development. CMII offers bachelor's and master's degree programs in game design, game development, media entrepreneurship, and digital filmmaking. The institute features production studios, audio and post-production suites, VR demo labs, and a VR "cave" with immersive projection (Creative Media Industries Institute, n.d.). The CETLOE Learning Design team includes learning experience designers, multimedia designers, and user experience researchers, who partner with faculty to create high-quality online, hybrid, and face-to-face learning experiences.

Leveraging our technological capabilities and deep understanding of curriculum development, we created Self and Community in the Metaverse, a course designed to test and evolve with the input of its participants and utilize a team-teaching approach. We taught with the XR tools of our trade and modeled their use in real-time, fostering an environment where students could learn by creating, critiquing, and iterating on their work through collaboration. This project used a student technology fee grant to provide Oculus headsets for students. However, it is important to note that most learning activities utilized WebVR, which did not require headsets.

## Participants

Twenty-four students enrolled in Self and Community in the Metaverse in the spring semester of 2023. The course was cross listed in the CMII Media Entrepreneurship B.I.S. program and the Entrepreneurship B.B.A. program housed within the Robinson College of Business. Prior to the class start date, we obtained an exemption from our institution's IRB to ensure that our process for collecting student feedback complied with ethics standards. We distributed a survey early in the semester to capture data about students' experiences with media and technology, their attitudes toward emerging tech, entrepreneurial tendencies, and demographic information. The intention of this survey was to gather data about our students to inform future design decisions. Of the sixteen survey participants, seven students majored in Media Entrepreneurship, with the remaining participants identifying as business, entrepreneurship, film, interdisciplinary studies, finance, game design, marketing, and psychology majors. Among the students participating in the survey, seven identified as male, eight as female, and one as non-binary/third gender. In terms of racial and ethnic background, the group included two individuals of Asian descent, seven who identified as Black or African-American, and six as White. Additionally, one participant did not identify with the racial and ethnic categories provided. Two individuals identified as Hispanic or Latino.

The first section of the survey was adapted from the Media and Technology Usage and Attitudes Scale (MTU-AS) (Rosen et al., 2013). The MTU-AS measures participant behavior and perceptions via a frequency scale indicating how often participants create or consume media on different devices and a Likert scale to measure attitude toward technology and society and the impact of the metaverse both personally and at large. Our data indicated that students' most frequent media-related activity was watching media, particularly on a computer or mobile phone. Almost all the students reported that they never or rarely used AI creation tools like Dalle-2 or mid-journey. Students took this survey in January 2023 at the beginning of the spring semester. The students' AI usage results reflect that moment in time and would likely change in future semesters.

Results from the attitudinal questions suggested that the students hesitated to embrace new technologies and exhibited concerns about AI co-creation. Students tended to feel connected to others through technology and acknowledged the importance of keeping up with technological trends. Some students also expressed ethical concerns regarding the capabilities of technology to create isolated bubbles and environments as well as its vulnerability to misuse.

Furthermore, the pre-survey data expressed the desire to understand how XR technologies could help them explore new opportunities, broaden their knowledge, and inform their prospective career pathways. Most students felt they were more likely to consume content online than create content and believed that technology enhances productivity. These student perspectives shaped the personas that formed the basis of our design process.

# Design Process

**Design Framework**. The design framework we used in designing the course Self and Community in the Metaverse was anchored within human-centered design thinking. Design thinking is a framework that encourages designers to learn about the people most impacted by design decisions. Design thinking is practiced through five iterative, non-linear actions: empathize, define, ideate, prototype, and test (Dam, 2024). Typically, designers empathize and define learner needs through interviews, focus groups, observations, and the development of personas. In this project, we also prototyped and tested a learning activity to inform design decisions.

**Personas**. Personas are fictional models that embody the needs, desires, preferences, and demographics of individuals participating in a designed experience. Designers utilize personas to ensure that the creation and development of an experience remain focused on the end user (Jerald, 2015). To inform our personas, we adapted survey instruments to measure learners' self-perceptions of their entrepreneurial competencies (Maldonado Briegas et al., 2021). Using the Technology Acceptance Model, we also examined how learners perceive tools in terms of enjoyment, ease of use, and usefulness to future goals (Lee et al., 2019). We included these instruments in the survey distributed to students at the beginning of the semester. After empathizing with student voices from the initial survey as well as interviews and focus groups, we defined learner needs through the following personas.

One student persona, the Media Entrepreneurship major interested in fashion design, is itching to build her own business in virtual worlds, where she's already tested tools, built skills, and created a community. Another persona wonders if XR environments will heighten isolation or if real-life (IRL) problems will be reproduced in virtual worlds. He perceives the relationship between emerging technologies and power: "I hope people harness the power for better and benefit, and don't take advantage of it, because the ones who know how to use it are going to be the ones with the power." The personas vary in prior knowledge, skepticism or enthusiasm for new technologies, and experience experimenting with XR tools. Some students have plunged into XR technologies, developing their expertise. Others do not know where to begin or if they even want to adopt such tools and platforms. These personas demonstrated the need to design flexible learning activities that could address varied student perspectives. During the course implementation, these needs guided our curation of support materials, low-stakes practice activities, and reflective pauses during workshops and other learning activities related to XR technologies. To meet the needs of our students, we also focused on course communities, such as Discord and in-class team projects, where students could share knowledge and perspectives.

**Learning Activity Prototype.** In addition to developing personas, we practiced human-centered design thinking by developing and testing a learning activity prototype, which allows designers to experiment with possible solutions to design problems and evaluate the efficacy of those solutions before implementing the course. We recognize that the structure of course development in higher education often short-circuits possibilities for design thinking, especially prototyping, testing, and iterating. Faculty are expected to prepare an entire course before interaction with students, and changes reflecting student experience only happen in future semesters. Given our commitment to the human-centered design thinking framework, we attempted a practical model for prototyping, testing, and iterating in the higher education context. To that end, we conducted a learning activity test before the full course began, in partnership with the Digital Learners to Leaders (DLL) internship program. DLL includes an experiential technology skill course that is co-sponsored by CMII and CETLOE, exposing student members to a broad spectrum of technologies. As guest lecturers in the DLL course, we conducted our avatar creation design sprint from Self and Community in the Metaverse.

A design sprint is a fast and structured process for solving problems and testing new ideas with end-users, in this case, students. Developed initially by Google, a sprint is typically divided into stages: understanding the problem, ideating solutions, deciding on the best solutions, prototyping, and testing (Knapp, 2016). Applying this approach to lesson planning enables teachers to rapidly refine teaching strategies, ensuring lessons effectively engage students. In this context, the avatar creation design sprint was a four-phase activity designed to equip students with the skills to create avatars for virtual environments. To scaffold the technological complexity of some avatar creation tools, students used pen and paper to create their first avatar. They then used ReadyPlayerMe and Metahuman. Finally, they navigated their avatars through a custom Spatial environment, with each phase followed by a reflective pause. The sprint not only introduced WebVR tools and environments but aimed to demystify the technology and empower students to realize their visions with these tools. UX researchers observed the class, deployed surveys, and conducted interviews for later review.

**Design Decision Rationale**. During the avatar design sprint activity test, we learned that avatar creation was not a low-stakes activity – regardless of grade points, there is nothing low stakes about asking underrepresented students to create representations of themselves. Barriers and limitations in XR technology, such as options for skin tone, hair style and texture, eyeglasses, body type, gender expression and other choices, can prevent students whose identities are underrepresented in technology sectors from seeing and creating themselves in XR. Students also experienced anxiety stemming from their lack of purpose when entering a virtual world.

While we couldn't prototype and test every learning activity, we applied these insights across course design. For example, we iterated the avatar creation activity from creating a self-representation to creating a character to navigate virtual worlds. While still inviting critical reflection on the limitations of the tools, we refrained from developing course activities that required self-representation. We also created tangible goals for each learning activity, such as completing a scavenger hunt when navigating a new virtual environment. One limitation of our design process was that we did not engage in focus groups, interviews, and surveys of learner needs until the avatar creation design sprint activity test. If opportunities exist, we recommend seeking student input and feedback on the goals and assessment strategy of the course, in addition to the learning activities.

The insight that students experienced anxiety from a lack of clear purpose with XR technologies impacted the development of core learning activities such as a live class session called Creating Digital Assets. We predicted that students would experience disorientation using 3D-modeling software, so we identified a simple task, the creation of a chair, to ground learner experience. Based on our personas, we expected a range of prior experience, so we curated preparatory materials that might meet varied needs, from an entry-level tour of 3D space to self-paced, adaptive tutorials for modeling software. We also invited an industry expert to facilitate the activity, to make the connection between course learning outcomes and the professional relevance of the more challenging tools.

Both the personas and the learning activity prototype highlighted the need for students to navigate, learn, use, compare and reflect on disruptive technologies as empowered creators. To facilitate such opportunities in our course design, we enacted problem-posing education through a capstone project that asked students to identify problems in their current university and build solutions in XR platforms (Freire et al., 2014). Rather than replicate the power structures that still trouble brick and mortar universities, we wanted to use the potential of 3D virtual worlds to create learning environments where the teacher is "taught in dialogue with the students, who in turn teach while also being taught" (Freire et al., 2014, p. 80). As students began to work on the capstone project, the instructor facilitated design thinking workshops using convergent and divergent thinking to gather perspectives, define problems, and ideate solutions. In teams, students then prototyped and iterated those solutions as XR environments. This step-by-step approach attempted to fuel curiosity and demystify fears of new content production pathways and technology. Emphasizing the constant nature of change, the course was structured to not only give knowledge about current technologies but also create adaptability and a mindset geared towards continuous learning.

Human-centered design thinking challenged us to listen and grow our work in response to learner experience. XR education, as well as education in other emerging technologies, offers an opportunity to decentralize the traditional hierarchies of education, and calls for curiosity, agility, and resilience from both educators and learners.

**Figure 1**

*Design Thinking For Learning*

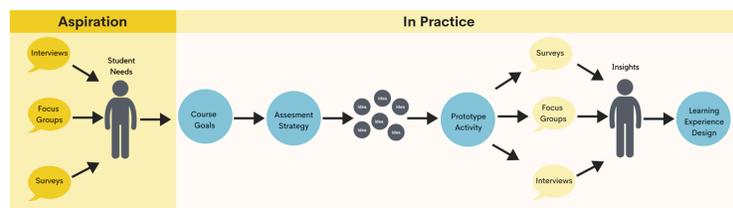

# Constraints

We have observed that gamified learning experiences, as well as personal consumption of XR, often position students as end-users of emerging technology. Assessing these end-user learning experiences is outside our scope, but this trend in XR learning proved challenging in our work to position students as co-creators. We found that most students only had end-user experience and faced barriers when learning technologies used for creation.

Consequently, learners in this course displayed varied experiences, skillsets, and motivations in relation to XR. Challenges include limited prior experience with mouse buttons and keyboard shortcuts, as well as limited off-campus access to computers with relevant software and computing power. Data from the focus group research sessions revealed that students grappled with a "learning curve" that created a barrier for those less experienced with the hardware. One student stated, "I feel like people who have a lot more experience with doing this on the computer had a lot easier time with this project." Consequently, navigating issues around digital literacy and accessibility emerged as a critical pain point within the learner experience and course design. These issues constrained our ability to design asynchronous learning activities or projects based on the level of in-person support needed. We also could not assume home access to software or compute power necessary for some XR technologies, so we focused our course on web-based XR.

These challenges were compounded by the lack of up-to-date and user-friendly documentation, tutorials, and onboarding materials for emerging tools since emerging technology is characterized by rapid cycles of iteration. The lack of up-to-date documentation and onboarding materials for emerging tools contributes to the alienation many students already feel about emerging technologies. In response, we designed learning activities to troubleshoot and test a variety of platforms and encouraged peer and instructor support in the course discord community. In addition, platform sustainability can be unpredictable. For example, one of the worldbuilding platforms we considered for the course, Altspace VR, had been decommissioned by the midterm of our teaching semester. Had we chosen this platform, students would not have been able to access their capstone projects for the final months of the course. In response to the volatility of emerging technology, we sometimes chose more established platforms despite their limited functionality.

An additional constraint is that learners report negative physical experiences using headsets (Taçgın, 2020), especially beyond a 15–30-minute time frame. In one pre-course interview, a student noted that prior experience wearing a headset "threw me off physically. I'd get headaches, and I was really weak."  While our participants did not include any students who disclosed motor or sensory disabilities that limited their participation, the lack of accessibility features in XR technologies continue to be an impediment to widespread adoption in learning environments (Ziker et al., 2021). Because of this constraint, we limited headset-based learning activities to short periods of time and included reflective pauses in headset use.

# Learner Experience

## Learning Environment

Self & Community in the Metaverse operated across multiple platforms throughout the semester. We first imagined a fully remote course with students having headsets at home, but quickly realized this would not be possible. Instead, the class met weekly in CMII's research lab space for interactive activities using a variety of metaverse creation tools. The classroom was equipped with PCs capable of running 3-D modeling, and browser related software. However, the assignments were intentionally designed to be executed on WebVR only, meaning a mobile phone would be sufficient for most of the work done in the class. The students also got access to Meta Quest 2 headsets for in-headset creation tools and to understand the look and feel of their virtual worlds in a spatialized setting, although these headsets were not a key tool in the class. The class also met on video conferencing platforms for additional technology exploration and group discussion.

A module-based course hosted on GSU's learning management system supplemented the in-class activities. The course served as a hub for students, offering topical resources such as curated readings and videos, supplementary materials, in addition to quizzes and surveys to check students' comprehension of key concepts. To encourage collaborative learning and foster a sense of community, the course utilized a dedicated Discord server for assigned discussions and general course communication.

## List of Platforms & Tools Used in Project

- **Midjourney:** an AI text-to-image generator - https://www.midjourney.com
- **ShapesXR:** a collaborative prototyping tool - https://www.shapesxr.com
- **Meta Quest 2:** a virtual reality headset - https://www.meta.com/quest/
- **MetaHuman:** an avatar creation tool - https://www.unrealengine.com/en-US/digital-humans
- **Ready Player Me:** an avatar creation tool - https://readyplayer.me
- **Spatial.io:** a virtual world platform and building tool - https://spatial.io
- **Hyperfy:** a virtual world platform and building tool - https://hyperfy.io
- **Blender:** an open-source 3D asset creation tool - https://www.blender.org
- **Sketchfab:** a 3D asset repository - https://sketchfab.com
- **Dalle-2:** an AI text-to-image generator - https://www.openai.com/dall-e-2
- **Discord:** a chat platform for community building - https://discord.com
- **Metaversity:** a virtual reality campus with educational experiences - https://metaversity.io

## Learning Activities

In this project, human-centered design thinking served as both design principle and course assessment strategy. The design thinking actions – empathize, define, ideate, prototype, and test – imply a continual practice of reflection, which course learning activities scaffolded more explicitly through reflective pauses, focus groups, and peer feedback. When it comes to applying human-centered design principles to actual entrepreneurial activities, the project integrated real-world tools like Shapes XR for prototyping and Blender and Sketchfab for creating or finding digital assets. This hands-on experience is more than just an academic exercise; it mirrors industry standards, equipping students with a deep understanding of new technologies. Mastery of these tools is vital as they reflect common practices in the field, enabling students to discover and create business efficiencies and innovations, particularly in the realms of spatial 3D creation and digital twinning. As students become proficient with these technologies and the broader virtual world ecosystem, they unlock new entrepreneurial opportunities. Because human-centered design thinking grounds problem-solving in the needs of the people most impacted by designer decisions, we have included learner perspectives within the descriptions of key learning activities.

**Learning Activity 1: Co-Create A Learning Environment.** In this activity, students began to prototype their final project using midjourney, then co-created 3D prototypes of these environments using Meta Quest headsets and the collaborative spatial design platform Shapes XR.

In surveys and focus groups, students shared positive experiences with the creative and collaborative aspects of this activity. "I like that it provided you with creative freedom," said one survey respondent. "It was cool to invite others into your space to create alongside each other." Students experienced challenges with the technology, including a new technology learning curve. In the focus group, some students did report transferring skills from other technological experiences when frustrated, "just like using a new phone." Other participants preferred the "camaraderie" of problem solving in a team, compared to the "silence" of previous activities: the team members were "right there next to each other, and having similar problems, like if you have a problem, we're just communicating what we're doing and interacting." The focus group mentioned fun, giggling, and laughter multiple times, and we also noted this heightened sense of play in our observations. While the surveys noted a range of student opinions on whether the benefits of the technology outweighed the effort, all respondents either planned to use or were considering using this technology in the future.

Some students did report that this activity caused physical discomfort, and students with eyeglasses were particularly impacted. One survey respondent shared that "vision wise it was not the best. Kind of gave me a headache." Our activity was structured to include pauses in headset use, and some students chose to engage with the technology for longer periods than others. In any case, XR learning must continue innovating to provide alternative means to access the valuable experiences of creation and collaboration that headsets can provide.

**Learning Activity 2: Creating Digital Assets**. To build their learning environments, students needed to create their own 3D assets. Before this learning activity, students completed self-paced video tutorials and tested the open-source 3D modeling tool Blender. In the live session, a professional 3D artist guest speaker led a workshop to create a simple environment and chair using the tool. We provided supporting resources such as a keyboard shortcut reference sheet.

In the focus group, some students shared that the expert-led workshop helped them persist through challenges with the technology. One student reported that "[in the tutorial], I just kind of got stuck and gave up because it just felt like there was too much I didn't know how to do." In contrast, when they worked with the expert guest speaker, "it was good to get in [the classroom] and someone knew." Students also found value in learning from someone who applies the tool in their own career. Other students reported that the workshop format tried their patience, because when individuals had challenges, the expert couldn't move on until the entire group was up to speed. Others valued how, unlike online tutorials, the workshop allowed them to see and learn from their peers. The surveys showed a range of student opinion on whether the benefit of the tool outweighed the effort, but all respondents agreed or somewhat agreed that the technology would help them with their career goals. Students shared positive perceptions of the open-source aspect: "I liked that the software was free and feels really accessible to use."  We also observed that students who completed the tutorial and gained prior familiarity with the software were more equipped to engage in the workshop and explore the tool while waiting for others.

# Discussion

The Self and Community in the Metaverse course exemplified how immersive technologies like XR can revolutionize educational practices at Georgia State University. Utilizing human-centered design thinking, transformative learning, and problem-posing education, the course enabled students to transition from passive consumers to active creators within 3D virtual worlds. The course demonstrated the relevance of XR technologies to varied career goals through projects and assignments linking theoretical knowledge with real-world applications. This approach was intended to help students perceive the practical implications of their learning and how these technologies could be applied across different industries. Reflecting on our design, we realized that our emphasis on hands-on workshops and peer-to-peer learning sometimes overshadowed the need for a more solid theoretical foundation supporting the 'why' behind the tools and activities. The rapidly evolving, collaborative, open-source ecosystem and critical questions of interoperability made it difficult to provide meaningful access and functional supportive resources for these tools and activities.

Moving forward, we aim to integrate more structured scaffolding to illustrate how these tools fit into various professional landscapes. This approach will ensure students not only know how to use the technologies but also understand their relevance across diverse career contexts. Our ongoing efforts will focus on enhancing how we present and integrate these complex systems, ensuring they are comprehensible and directly applicable to students' future professional environments. This entrepreneurial approach explores the intersection between academic and industry practices. As XR technologies continue to evolve, they present opportunities to enhance instructional practices and outcomes, emphasizing the importance of agility in adopting new educational technologies alongside maintaining rigorous academic standards.

Looking ahead, it is necessary for academic institutions to remain at the forefront of integrating emerging technologies. Continuous exploration of XR's pedagogical potential, coupled with a commitment to addressing accessibility issues, will be crucial. To this end, we have formed a partnership with the Immersive Experience Alliance (https://ixalliance.org/), a coalition of public universities focused on XR programs that merge physical and digital spaces to create unique and novel experiences. This collaboration aims to bridge gaps among NGO leaders, academic institutions, and immersive media creators, fostering learning, networking, and identifying opportunities for potential collaborations. While the Self & Community in the Metaverse project showcases the need for such partnerships and collaborations, the implications of our design also provide insight into how instructional design principles such as human-centered design, transformative learning, and problem-posing education might need to be reconsidered in the context of disruptive technologies.

# Implications

## Decentralized Pedagogy

The ethos of Web3 is decentralized and peer-to-peer. This non-hierarchical trend echoes the combination of instructional design principles we found most effective in this project, which, in their combined use, facilitate a form of decentralized pedagogy. While problem-posing education's call to restructure the teacher-student relationship (Freire et al., 2014) is an important starting point, learning experiences cannot effectively position students as creators or emerging experts without a human-centered approach to identifying their needs. Static subject matter expertise is not agile enough to empower students in a dynamic landscape, so in this project, effective XR learning required partnership between instructional designers, user experience researchers, media designers, 3D artists, and metaverse practitioners. When possible, we practiced transparency regarding the collaborative ecosystem of this design, with instructional and multimedia designers engaging directly with students. These layers of collaboration and co-learning positioned instructors to facilitate rather than transmit expert knowledge. To effectively apply Freire's model to the realities of disruptive technologies as subject matter, universities and departments need to develop intentional pathways for collaborative instructional design.

**Figure 2**

*Ecosystem Map of Decentralized Pedagogy*

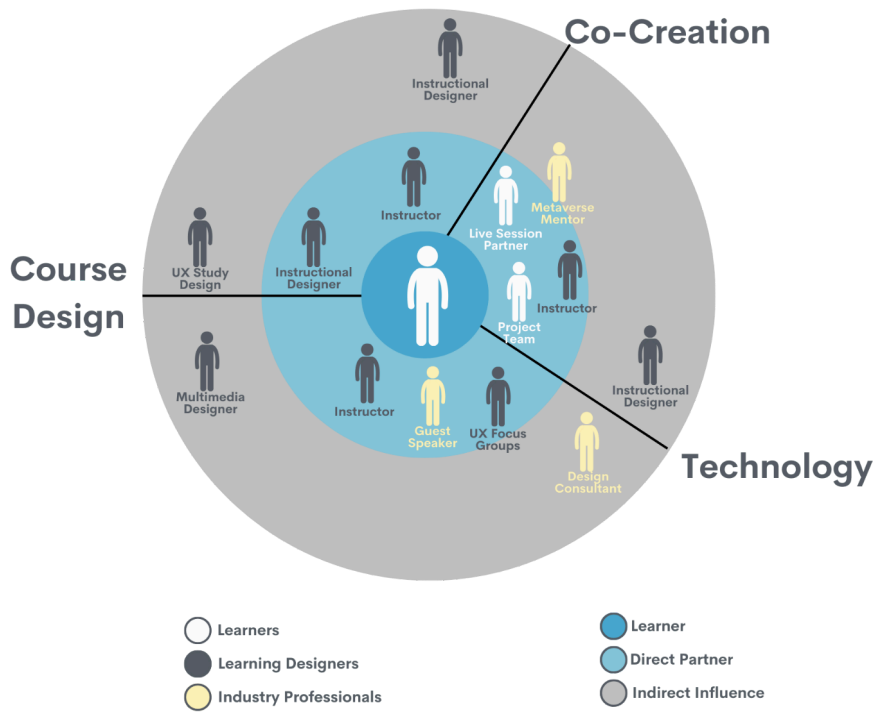

# Implications of the Student Experience

Throughout the course, students expressed uncertainty about the purpose of learning activities, often in activities designed to create agency by exploring, comparing, choosing, and creating across a set of tools. In one such creative activity, students reported, "it took our group a little while to get started on like, what we should do...you know, that would have kind of given us a little more direction to start out with." The discomfort described by the students also aligns with "disorienting events" in transformative learning. These disorienting events occur when the student is presented with a situation that directly conflicts with their prior experiences. Disorienting events stimulate critical reflection and cause the student to question their assumptions of social norms (Cranton, 2016).

Student reports of confusion and dislocation expose the degree to which students have been socialized through prior educational experience to receive rather than create knowledge. Similarly, our decentralized approach exposes the lack of agency and creative positioning in previous student experiences with technology. Although transformative learning accounts for disorienting events, this project suggests that such events can impede student learning when paired with the disorientation of new technologies. We suggest that the following practices can mitigate the negative consequences of these destabilizations.

## Key Practices for Decentralized Pedagogy

- Get comfortable with discomfort. Instructors should be honest about their own confusions and model navigating these confusions for students.
- Embed reflective pauses in course design to facilitate growth and help students process disorientation (Baldwin et al., 2020).
- Model curiosity and investigate where experiences of messiness and disorientation might come from.
- Provide clear rationale and support resources when embedding industry practices in undergraduate education.
- At the institutional level, provide structures that empower collaborative course design and instruction.

## Assessing & Learning XR Technologies

Another complicating factor for the application of problem-posing education and transformative learning to XR education arose from the rate of change in the subject matter. During our year of design, teaching, and research, we witnessed major changes to the most available, relevant, and usable XR technologies. We propose the following method in Figure 3 for assessing and learning disruptive technologies of any kind. Our framework might be used by instructors and course designers to plan, but also by students to compare and learn different tools as a recurring learning process. Rather than focusing on the tools themselves, this approach empowers both instructors and learners to evaluate, choose, and create with technology. For entrepreneurs, the ability to identify and skillfully utilize the right tools effectively and efficiently stands out as a distinct advantage, distinguishing them from the competition. Knowing how to choose the right tool is a key component often missing from the technological literacy necessary for success across many contemporary industries and disciplines.

**Figure 3**

*Framework For Selecting Platforms & Tools*

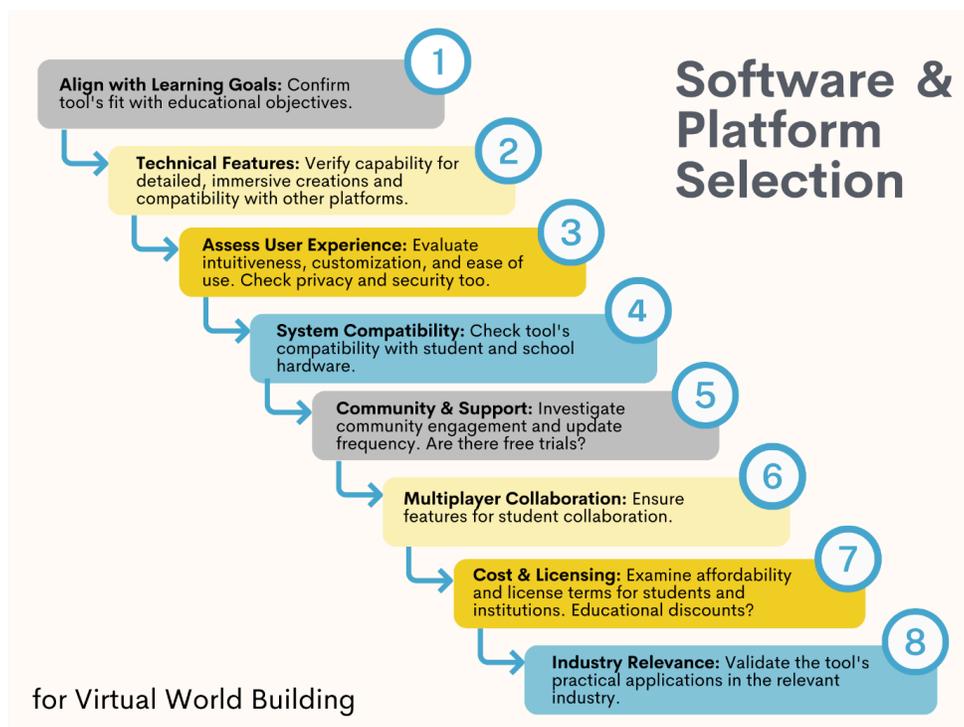

# Conclusion

The human-centered course design and implementation described in this practice paper offer a blueprint for any learning experiences impacted by disruptive technologies. XR's dynamic business opportunities make Media Entrepreneurship a productive use case for human-centered design, transformative learning, and problem-posing education, but students across many academic disciplines also need learning experiences that support them on the unstable ground of rapid technological change. Human-centered design provides an effective framework for how universities might define and respond to such needs, but this adoption will require new processes and structures so that collaborations between user experience researchers, designers, and instructors can be valued and scaled. In advance of such structural changes, transformative learning and problem-posing education are useful approaches to change mindsets and empower students to create and produce with emerging technologies. However, our experience also highlights how the disorientation inherent in these instructional principles is compounded by the disorientation experienced by users of emerging technologies. When practicing transformative learning and problem-posing education, designs need to include clear goals, collaborative learning, empathetic guidance, and robust reflective scaffolding. These practices enable learning experiences that continue with the important goals of destabilizing structures of power in both learning and technological experiences while also positioning students for success in professional futures impacted by disruptive technologies.

# Acknowledgment

 The authors wish to acknowledge Morgan Nixon, Jackie Slaton, Chad Marchong, Maya Wilson, Garrett DeHart, Liv Cambern, Stu Richards, Angel VanEllison, James Amman & Robert Paraguassu for their contributions to the design, implementation, and research of Self & Community In The Metaverse.